# Evidence of liquid-liquid phase transition in compressed Ar probed by the thermal expansion of Mo, Ta and W at high pressures

Joseph Gal*

Ilse Katz Institute for Nanoscale Science and Technology ,
Ben-Gurion University of the Negev, Beer Sheva ,84105 Israel

## Abstract

The long standing controversy between the melting curves of the bcc Mo,Ta,W and vanadium (V) metals measured by diamond anvil Cells (DAC) and the shock dynamic experiments is explained by the behavior the liquid or solid pressure transmitting mediums compressed by the thermal expansion of the these transition metals. This explains the observed isobaric behavior of the laser heated DAC experiments containing different transmitting mediums reported in the literature [1, 2, 3], thus solving the standing enigma described in very many publications.
It is shown that the Ar pressure transmitting medium is in its liquid state which is compressed by the thermal expansion of the Mo,Ta,W metals. The liquid Ar shows evidence of a liquid-liquid phase transition. This is observed by a density (viscosity) change upon increasing the pressure and the temperature in the laser heated DAC. Two density regions are detected: below ~ 40GPa where $dT_m/dP >0$ and above ~40 GPa where $dT_m/dP \rightarrow 0$. The different slopes of the melting curves are enlarged in the insets of Fig.s 2-4 indicating different resistance to the thermal expansion of the transition metals relative to the softer liquid Ar. The melting anomaly observed above 40GPa is discussed.

*Introduction*-The standing controversy between the static DAC measurements and the dynamic shock wave melting curves of the bcc transition metals is an important physical issue that attracted the high pressure researchers for long time. It is about two decades since D. Errandonea, B.Schwager, R. Ditz, C. Gessmann, R. Boehler and M. Ross published in PRB the paper " Systematics of the transition metals melting" [1]. The high-pressure-temperature measurements of Mo,Ta,W and V metals were performed in a laser-heated diamond anvil cell (DAC) up to nearly 100 GPa and 4000 K where most data were obtained in argon pressure transmitting medium (PTM). Small melting slopes which approach zero at high pressures were reported hinting isobaric conditions in the DAC. All the experimental dynamic shock waves data (SW) and first principle ab-initio DFT calculations for the melting of these transition metals are incompatible with Errandonea's et al. DAC experiments. Nevertheless, no conclusive explanations of this behavior can be found in the literature, thus up to date the discrepancy between the SW data and the DAC data remains enigmatic.

It has been shown by Dewaele et al.[2] and recently be Errandonea et al. [3] that this isobaric behavior is strongly related to the behavior of the PTM used in the DAC experiments. Commonly used for pressure transmitting mediums are Ar, $Al_2O_3$, KCl , NaCl, MgO [3]. In the present contribution the behavior of Ar PTM is analyzed taking into account the negligible chemical reaction of the medium with the investigated sample and with the compressing diamonds [1].

Studies of compressed noble gases revealed different slope of the melting curves of Ar,Xe,Kr which were related to the fcc and the hcp crystallographic structures [4]. Recently, the extrapolated melting curve of compressed Ar, utilizing Gilvarry-Lindemann approximated (LG) criterion, indicated that under pressures and above 4500K, Ar is absolutely in the liquid state [5] . In the present paper it is claimed that the experimental data measured by D. Errandonea et al. shows evidence of a liquid-liquid phase transition in Ar which is observed by a density (viscosity) change upon increasing the pressure and the temperature in the laser heated DAC.

In conclusion, it is claimed that the discrepancy between the SW data and the DAC data or the ab-initio DFT calculations [6,7,8,9] can be explained by the thermal expansion of the hard transition metals Mo, Ta, W pressing the softer Ar liquid PTM. This applies also to vanadium metal using different pressure mediums (see discussion). The enigma mentioned above is thus solved.

*The combined approach-* In previous publications it was shown that by introducing a constraint demanding that the fitting of the experimental equation of state (EOS) will simultaneously fit the experimental melting data. By combining the LG criterion with the EOS a consistent bulk moduli are obtained allowing the extrapolation of the fitted melting curves to high pressures and temperatures. Assuming harmonic Debye solids and adopting the LG approximation [10], good fits of the melting curve at high pressures and temperatures for Al,Cu and U metals as well as for Ar,Xe Kr, were obtained [5]. Utilizing the present combined approach and assuming negligible thermal pressure the melting curve and the EOS of compressed Ar were simultaneously fitted and parametrized as depicted in Fig.1:

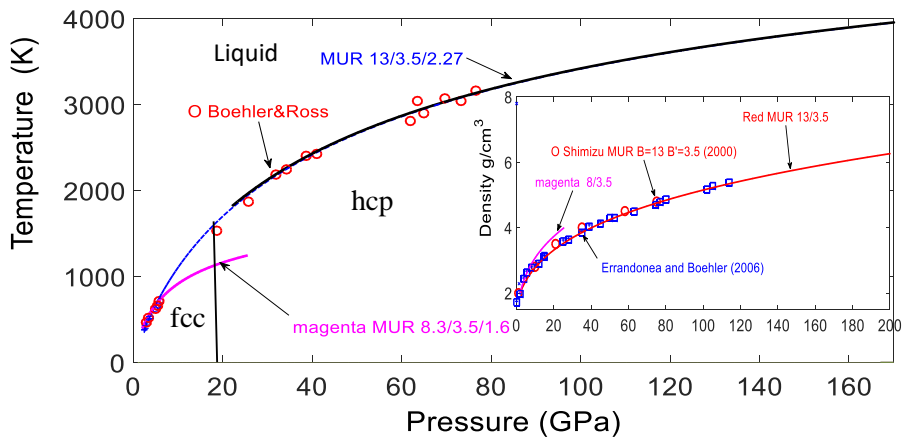

Fig. 1: **Argon** phase diagram and EOS. Melting curve of the fcc and hcp phases fitted separately for each crystallographic phase (fcc,hcp) utilizing the combined approach step 1 [5]. The EOS (P-V space) is shown in the inset. Note that the phase transition (slope) observed in the melting curve is also clearly pronounced in the EOS inset. The solid lines represent fittings appllying Murnaghan equation of state.

The solid lines are the best fits parameterized with $B_o$=8.5(2)GPa , $B_o$'=3.5 and $\gamma_o$ =1.6 (marked 8.5/3.5/1.7) for the fcc phase and $B_o$=13(2)GPa, $B_o$'=3.5 and $\gamma_o$ =2.10 (marked 13/3.5/2.10) for the hcp crystallographic region. The derived bulk modulus indicate that hcp-Ar is an extremely soft solid. Most important result is that in any pressure above 4500K Ar is in its liquid state.

*Molybdenum*- At ambient temperature and approaching the melt Mo is a body-centered-cubic (bcc-Im3m) 4d transition metal. The pressure dependent melting temperatures of the Mo metal is taken from D. Errandonea et al. [1]. These high-pressure measurements were performed in a laser-heated DAC with argon (PTM) up to nearly 100 GPa and 4000 K. Most striking in Ref.1(Fig.1) is that after melting no chemical reaction of the Mo metal with the tungsten (diamond coated) gasket or with the pressing diamonds can be identified. In addition, above 35GPa the slope approach zero indicating isobaric conditions in the DAC. Nevertheless, all the experimental dynamic shock waves (SW) data [7,14] and first principle ab-initio DFT calculations reveal incompatible melting curve with Errandonea's et al.[1] DAC experiments.

The melting of Mo metal measured by dynamic and static experiments are depicted in Fig.2:

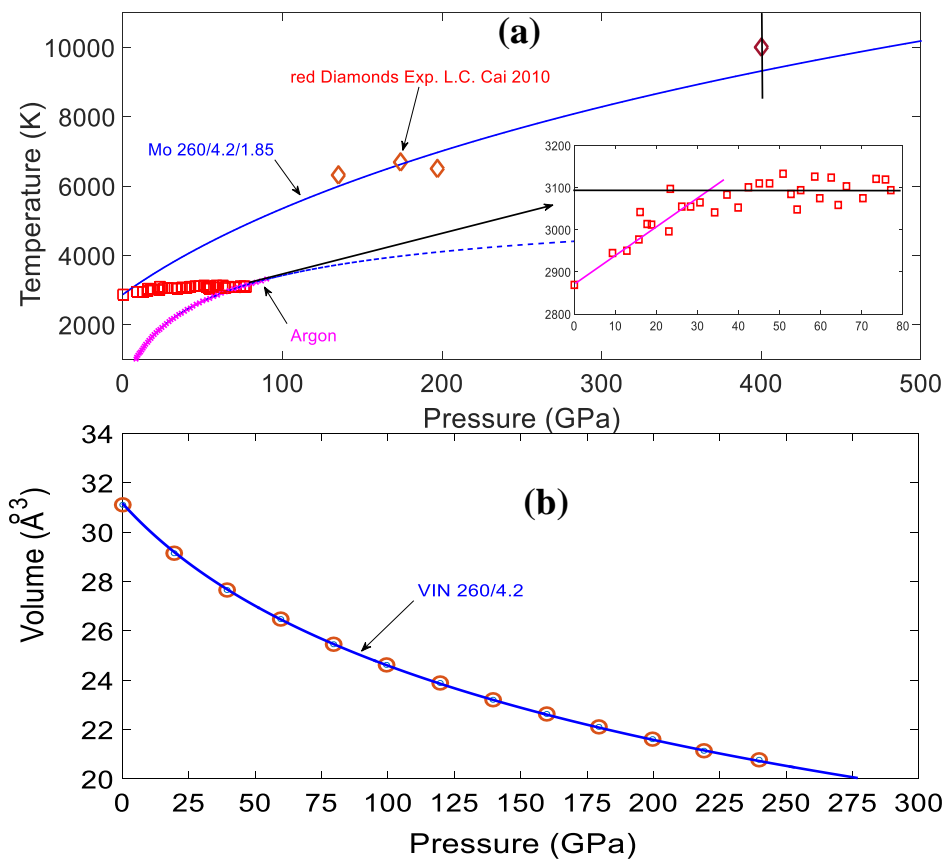

**Fig 2**: **(a)** Phase diagram of molybdenum metal. The red diamonds and purple square are the experimental SW data reported by L.C.Cai [7]. The solid blue line is the melting curve fitted with Gilvarry-Lindemann approximated criterion combined with Vinet EOS. The red squares are the melting points measured in a DAC with Ar PTM [1]. The inset clearly demonstrate two regions of melting slopes which are related to the melted Ar compressed by the Mo thermal expansion. The magenta x line and blue dashed line represent Ar melting curve. **(b)** Isotherm 300K fitted with Vinet EOS parametrized with Bo=260(4) and Bo'=4.2(2) blue solid line. The experimental measurements of the P-V data are averaged according to [6].

In Fig.2(a) the parametrization of the bulk moduli and the melting curve yield $B_o$=260(4) GPa and $B_o'$=4.2(2) and $\gamma_o$=1.85(1) (marked 260/4.2/1.85). The fitting and the extrapolation takes into account only the experimental SW melting data as recorded by L.C. Cai et al. [7]. By applying the LG criterion together with the EOS constraint (the combined approach) the extrapolated melting curve of Mo was obtained assuming that the SW data is the anchor. The magenta x line and the extrapolated dashed blue line are the Ar melting curve indicating that Ar is a liquid PTM. The enlarged P-T melting data in the inset clearly shows two different slope regions $dT_m/dP$=7K/GPa) and $dT_m/dP$=0K/GPa. In Fig.2(b) the blue solid line represents the 300K isotherm fitted with VIN EOS of the experimental P-V deta [6]. As the Ar PTM is in the liquid state the volume thermal expansion of the molybdenum metal compress the liquid Ar which show different densities in the two pressure regions, evidence of liquid-liquid phase transition.

Tantalum: Ta is a body- centered-cubic (bcc-Im3m) at room temperature and approaching the melt [15,16]. The pressure dependent melting temperatures of Ta metal confined in the DAC [1] were performed in argon PTM and a laser-heated DAC up to nearly 100 GPa and 4000 K. Above 35GPa the melting slope approach zero and definitely points to isobaric conditions in the DAC. Similar to Mo, all the experimental dynamic shock waves data (SW) and first principle ab-initio calculations [16,17] are incompatible with Errandonea's et al. DAC experiments [1]. Certain corrections using non equilibrium heat flashing [17] or different PTM [3] failed to match the melting curve of Ta based on the SW anchor.

The phase diagram and EOS of Ta, utilizing the combined approach are depicted in Fig.3:

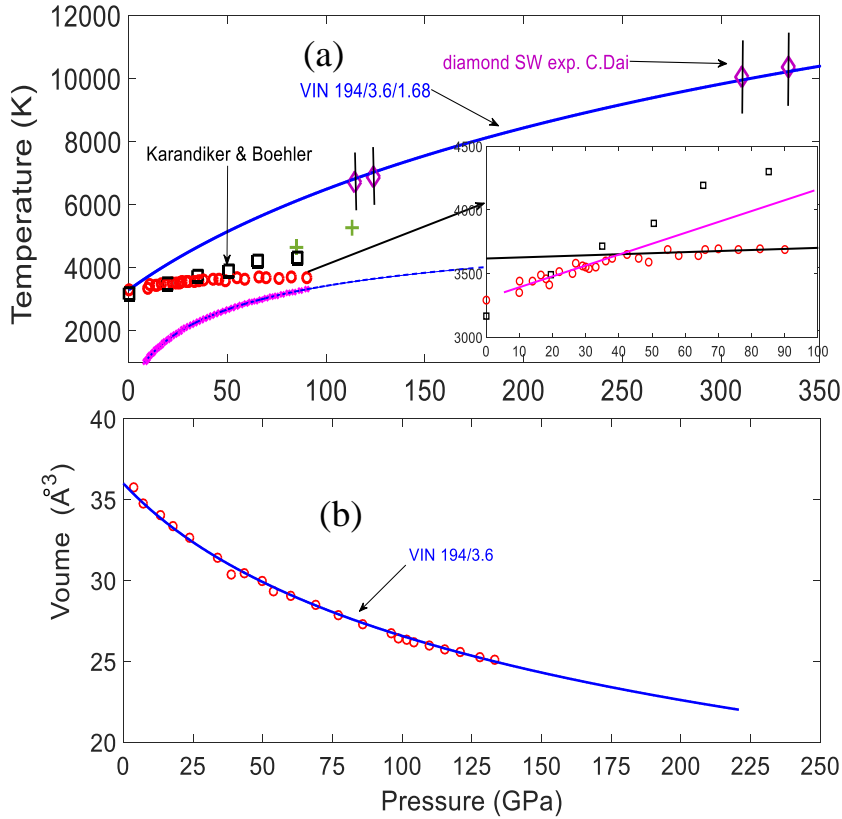

**Fig 3**: **(a)** Phase diagram of Ta metal. The purple diamonds are the experimental SW data reported by C.Dai [14]. The solid blue line is the melting curve fitted by the Gilvarry-Lindemann approximation combined with Vinet EOS constraint. The red circles, black squares and green plus are the melting points measured in DAC with Ar PTM reported by ref.s [1][2][17] . The magenta x line and blue dashed line represent Ar melting curve. The inset clearly demonstrates two regions of melting slopes which are related to the compressed liquid Ar. **(b)** Isotherm 300K as reported by H. Cynn et al. data [18], fitted with Vinet EOS parametrized with Bo=194(4) and Bo'=3.6(2), blue solid line.

The parametrization of the bulk moduli and the melting curve yield $B_o$=194(4) GPa and $B_o'$=3.6(2) and $\gamma_o$=1.68(1) marked 194/3.6/1.68 in the figure. The fitting and the extrapolation takes into account only the measured experimental SW melting data as reported by C. Dai at al.[14] and Haskins and Moriarty [16]. By applying the LG criterion together with the EOS constraint (the combined approach) the melting curve of Ta is obtained. The enlarged P-T melting data measured in the DAC, shown in the inset, clearly demonstrate two different regions of slopes, 0-40 GPa with dTm/dP=9K/GPa and dTm/dP=1K/GPa above 40GPa. As the Ar PTM is in the liquid state, the thermal expansion of the hard solid Ta metal compress the soft Ar liquid. Two different slop regions are clearly observed in the inset of Fig.3, reliant to the resistance of the Ar liquid, evidence of liquid-liquid phase transition in the molten Ar.

*Tungsten-* Approaching the melt W exhibits a body- centered-cubic (bcc-Im3m) crystallographic structure. The pressure dependent melting temperatures of the W metal confined in a DAC embedded in Ar PTM, as reported by D. Errandonea et al.[1], is shown in Fig 3(a). The measurements were performed under similar conditions of Mo and Ta. The experimental dynamic SW melting point is the only data reported in the literature [6,8]. The melting curve is constructed with the LG and VIN EOS (combined approach) and parametrized with Bo=300(4)GPa, Bo'=4.2(2) and $\gamma_o$=1.97 (300/4.2/1.97), in accord K. D. Litasov et al.[22].

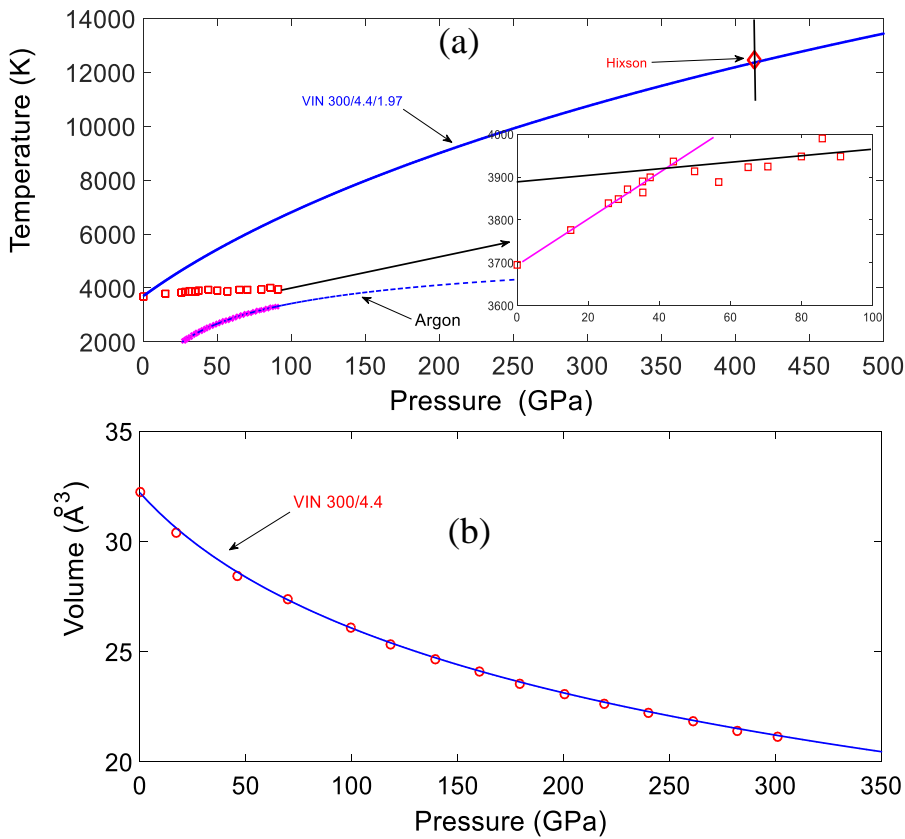

Fig.4 **(a)**: Melting curve of tungsten. The red diamond is the experimental SW data reported by Hixson et al. [6]. The solid blue line is the melting curve fitted by the Gilvarry-Lindemann approximation combined with Vinet EOS constraint. The red squares are the melting points measured in DAC with Ar PTM reported by D. Errandonea [1]. In the inset (enlarged) demonstration of the two regions of melting slopes which are related to the compressed liquid Ar. The magenta x line and blue dashed line represent Ar melting curve. **(b)** Isotherm 300K as reported by Hixson et al. data [6], fitted with Vinet EOS parametrized with Bo=300(4) and Bo'=4.4(2), blue solid line.

The enlarged P-T melting data in the inset clearly show two different regions of slopes, 0-42 GPa with $dT_m/dP=5.47$ K/GPa, and above 42 GPa $dT_m/dP \sim 1$ K/GPa. As the Ar PTM is in the liquid state (magenta X and dashed blue) the volume thermal expansion of the hard solid W metal presses the soft liquid Ar. The two different density regions are clearly observed, evidence of liquid-liquid phase transition in liquid Ar.

Discussion – There are three important contributions on the melting properties of Mo, Ta and W metals studied separately by laser-heated DAC utilizing Ar as PTM; the first is Arrandonea et al. [1] (2001) followed by C. Dewaele et al. (2010) [2] and A. Karandiker and R. Boehler [17] (2016). Each of the reported measurements yielded different melting, raising the question whether the interpretation of the experimental results are misleading. The latest paper [17] calls for a reevaluation of theoretical, shock compression, and diamond cell approaches to determine the melting at high pressures. I would rather say that the DFT simulations including the quantum molecular dynamics (QMD) predictions are, within the errors, in accord with the experimental SW data (see Fig.1,[17]). As will be discussed below the reason for the discrepancies comes from the thermal expansion interplay between the investigated sample and the PTM in DAC measurements.

There exist three situations in DAC experiments;

1. The melting temperature of PTM and the bulk modulus at ambient temperature are much higher than the investigated sample. For example, $Al_2O_3$, KCL or Ar are used in DAC experiments deriving melting points of Al and Cu metals. Here, at each applied pressure the thermal pressure increases upon raising the temperature in association with increase of the melting temperature, thus isochoric condition in the DAC exists [10].

2. The melting temperature of PTM is very much lower than the melting temperature and bulk modulus of the investigated sample. Namely, high compressible PTM as compared to extremely low compressibility of Mo,Ta,W and V, exhibiting very high melting temperatures. For example, liquids Ar, $Al_2O_3$, KCL or NaCl PTMs as compared to the above transition metals. Here, very small thermal pressure can develop leading to observed isobaric like situation in the DAC.

3. No PTM, meaning that the investigated sample (pressed powder) fills the whole volume in the DAC up to the gasket, preventing uniaxial stress, like for example in the case of Zr metal [19].

In the present study I concentrate on transition metals confined in a DAC imbedded in an Ar bath. Upon reaching the melt no chemical reaction with the diamonds with Ar or with the gasket has been observed [1,17]. The transition metals melting temperatures at zero pressure and ambient conditions are 2869K(Mo), 3290K(Ta),3655(W) (handbook). The bulk moduli as calculate above revealed high $B_o$= 260,194,300 GPa respectively, compared to compressed solid Ar $B_o$=8-13GPa and melting temperatures of 400-2200K, which upon raising the temperature all the way Ar is liquid.

The only experiment in which the samples were indeed completely molten is Errandonea et al. [1]. Starting from ambient pressure and temperature, the samples confined in the DAC sense small thermal pressure yielding increase of the melting temperature in the region 0~40GPa. Above 40GPa isobaric behavior is observed Fig.s 2-4, all in thermodynamically equilibrium state. Critics on the measurements of Ref.2 can be found in Ref.17 pointing that the measurements are performed in a non-equilibrium conditions. However, the measurements reported by A.Karandiker and R. Boehler [17] are completely different from the measurements of Errandonea et al. because of the following reasons: the flushing laser-speckle pattern on the sample surface forms a non-equilibrium heat spike in which the molten Ta expansion is suppressed by surrounding solid Ta metal. Thus, developing thermal pressure yielding higher pressure and elevated melting temperatures as depicted in Fig.2 black squares. In addition, the determination of melting by in-situ XRD is partially correct as this information relate only to the surface of the sample.

The recent (Sep. 2019) contribution of D.Errandonea et al [2] demonstrates that the melting curve and phase diagram of vanadium under high-pressure and high-temperature conditions measured by DAC, show closer matching to the SW experimental and theoretical DFT(QMD) calculations. They used NaCl or MgO as PTM which are mostly solid [21] and exhibits higher resistance to the expansion of the V metal, yielding thermal pressure which elevates the melting temperature accordingly. Thus, better matching the SW up to ~ 60GPa was obtained. Nevertheless, the pressure reported in Fig.5 of Ref.2 is not the actual pressure.

Finally, Errandonea et al. measurements were based on observation of a change in the emissivity of the surface until a complete melting was achieved. In addition, they used diamond-coated tungsten gaskets which are known for excellent sealing. It could be claimed that for some reason these gaskets leaks above 40GPa and the liquid-liquid transition observed is just an artefact. However, Errandonea at al. [1,21] reported that the in situ laser spackle method gave the same results where no changes in the surface texture on recovered or quenched samples were ovserved. Nevertheless, what is missing in their research are several pressure-temperature cycles looking for hysteresis effects that would prove that no Ar was lost. Trusting the high quality experimental work performed by D. Errandonea and his coworkers I conclude that liquid to liquid phase transition do occur in liquid argon under pressures and high temperatures. The fact that above 40GPa $Tm/dP \rightarrow 0$ is an anomaly that should be further considered.

It is proposed to re-measure the melting temperatures of Mo,Ta,W and V without PTM according to situation 3 explained above [19], predicting no discrepancy between DAC and SW results.

*Conclusions* – Thermal expansion of Mo,Ta,W metals imbedded in an Ar bath confined in a DAC compress the softer liquid Ar PTM. Two viscus regions are observed: 0 to ~40 GPa and above 40 GPa, evidence of liquid-liquid phase transition.

The DFT simulations including the quantum molecular dynamics (QMD) predictions are, within the errors, are in accord with the experimental SW data. The reason for the discrepancies between the SW and DAC experiments comes from the interplay between the thermal expansion of Mo, Ta, W, V samples vs. the soft PTMs in DAC experiments.

**Acknowledgement**

The author gratefully acknowledge Prof. Z. Zinamon, Department of Particle Physics, Weizmann Institute of Science, Rehovot – Israel, for the many helpful illuminating discussions and comments.